\documentclass{andp2012}
\usepackage[english]{babel}
\usepackage{amssymb,amsfonts}
\usepackage{graphicx}
\usepackage[caption=false]{subfig}

\def\sc{0.99} 

\def\art{Letter}

\def\jrn#1#2#3#4#5#6{#3 \textbf{#4}, #5 (#6).} \def\andd{and } 

\def\bfl{\begin{flushleft}}
\def\efl{\end{flushleft}}
\def\bfr{\begin{flushright}}
\def\efr{\end{flushright}}
\def\bc{\begin{center}}
\def\ec{\end{center}}
\def\be{\begin{equation}}
\def\ee{\end{equation}}
\def\bse{\begin{subequations}}
\def\ese{\end{subequations}}
\def\ba{\begin{eqnarray}}
\def\ea{\end{eqnarray}}
\def\baa#1{\begin{array}{#1}}
\def\eaa{\end{array}}
\def\bw{\begin{widetext}}
\def\ew{\end{widetext}}
\def\nn{\nonumber }
\def\lb#1{\label{#1}}
\def\bit{\begin{itemize}}
\def\eit{\end{itemize}}
\def\bco{}
\def\bcs{\begin{cases}}
\def\ecs{\end{cases}}

\def\twomat#1#2#3#4{\begin{pmatrix} #1 & #2 \\ #3 & #4 \end{pmatrix}}
\def\twomatsm#1#2#3#4{\left(\begin{smallmatrix} #1 & #2 \\ #3 & #4 \end{smallmatrix}\right)}

\def\fourcol#1#2#3#4{\begin{pmatrix}#1\\#2\\#3\\#4\end{pmatrix}}

\def\Sign#1{\, \text{sign}\left(#1\right) }

\def\densnnorm{\hat{\rho}^\prime}

\def\om{\Omega}
\def\ps{\text{ ps}}
\def\ips{\ps^{-1}}

\def\ave{\overline{E}}
\def\popdo{P_d}

\def\t{t}

\def\hamto{\hat{H}}

\def\av#1{\langle #1 \rangle}
\def\avo#1{\av{#1}'}
\def\bra#1{\langle #1 |}
\def\ket#1{| #1 \rangle}

\setcopyrightyear{2017}%
\DOIprefix{10.1002}%
\DOIsuffix{andp.201600185}%
\Volume{529}%
\Year{2017}

\date{224}

\category{
Rapid Research Letter
}


\title{
Sustainability of
environment-assisted energy transfer in 
quantum photobiological  
complexes
}

\subtitle{{
\small \rm Ann. Phys. (Berlin) \textbf{529}, 1600185 (2017) 
\qquad\quad DOI: 10.1002/andp.201600185
}}

\author[Konstantin G. Zloshchastiev]{Konstantin G. Zloshchastiev 
}
\address{K. G. Zloshchastiev\\
Institute of Systems Science\\
Durban University of Technology\\
P.O. Box 1334, 
Durban, 4000 South Africa\\
E-address:~http://bit.do/kgz
}

\shortauthors{K. G. Zloshchastiev}

\Receiveddate{: ~ 22 June 2016 [APS], 7 July 2016 [AdP], 13 April 2018 [arXiv]}

\shortabstract

\begin{abstract} 
It is shown that quantum sustainability is a universal phenomenon
which emerges during environment-assisted electronic excitation energy transfer (EET) 
in photobiological complexes (PBCs), such as photosynthetic reaction centers and centers of melanogenesis. 
We demonstrate that quantum photobiological systems must be sustainable for them to simultaneously endure continuous energy transfer and keep their internal structure from destruction or critical instability. 
These quantum effects occur due to the interaction of PBCs with their environment which
can be described by means of the reduced density operator and
effective non-Hermitian Hamiltonian (NH). 
Sustainable NH models of EET predict the coherence beats, followed by the decrease of coherence down to a small, 
yet non-zero value. 
This indicates that in sustainable PBCs, quantum effects survive on a much larger time scale than the energy relaxation of an exciton. We show that sustainable evolution significantly lowers the entropy of PBCs and improves the speed and capacity of EET.
\end{abstract}

\begin{document}
\maketitle


Typically the effect of solar radiation on living organisms and organelles begins with the absorption of a sunlight photon by pigments, followed by transfer of its energy to the reaction center, where primary electron transfer reactions transform solar energy into electrochemical gradient. One example of this process would be the natural photosynthetic stages, such as the Fenna-Matthews-Oslov 
complexes, which exist in green sulfur bacteria and marine algae. These complexes are parts of
complex biochemical structures that capture quanta of visible light into their peripheral light-harvesting complexes
and funnel the excited energy to the photochemical reaction centers, where it is used to initiate chemical reactions \cite{bm75,lzb97,wgr02,xsg04,bsv05}, see also monograph \cite{blbook} 
and references therein.
The efficiency of this transfer is very high, the reason for which has not yet been fully understood. Another example, closely related to human physiology, is the formation of cyclobutane pyrimidine dimers
caused by photons of the ultraviolet part of the sunlight spectrum \cite{fhg89,wpc01}.
This results in direct DNA damage, and triggers the process of melanogenesis: a synthesis of the melanin pigment contained in a special organelle called a melanosome \cite{ay05}. 
The duration of the chemiexcitation of melanin derivatives after ultraviolet exposure is reported to be unexpectedly long \cite{pwm15},
the reason for which has also not yet been completely revealed. In this \art, we propose a universal mechanism that explains the long-life and high-efficiency phenomena happening in photobiological complexes, and we also suggest analytical tools for their quantitative description.

In spite of having different chemical structures and spectra of absorbed light, all photobiological systems share certain universal features. For instance, they all operate in a thermal environment at physiological temperatures, and in the presence of external sources of noise and dissipation. Therefore one would expect quantum effects to be negligible. 
However, experiments based on two-dimensional laser-pulse femtosecond photon echo spectroscopy 
reveal the long-lived exciton-electron quantum coherence in photosynthetic reaction centers of different organisms \cite{sbs97,ecr07,lcf07,ran14}. 
Taking this into account, EET in PBCs can be described by applying existing quantum-mechanical 
methods \cite{cs84,fbook}. 
Numerous studies have demonstrated that environment 
plays significant role in energy transfer in light-absorbing complexes  \cite{le96,pp01,mrl08,oc08,ph08,if09,rmk09,cdc10,nbb12,nmm15,mmn15}.

Moreover, it is believed that it is the very presence of such a dissipative environment that increases the efficiency of EET to such a high degree. While this seems somewhat counter-intuitive from a classical point of view, it can be explained on quantum-mechanical grounds. In the absence of a thermal environment, the excitonic-type EET dynamics in photochemical reaction centers is dominated by coherent hopping, therefore the system is largely disordered and exhibits Anderson localization \cite{an58}.
Under such conditions, localization functions as the energy conservation mechanism of excitonic states: an exciton originally localized at an initial site is a superposition of energy eigenstates, which has only a slight overlap with an excitonic state which is localized at a final state.  Therefore, strong coherence would lead to a low efficiency of EET from one site to another. 
Once thermal effects come into play, the coherence starts to be destroyed. If the magnitude of dephasing effects is only just sufficient to destroy coherence-caused localization, the excitations are set ``free'', and consequently the efficiency of EET will rise drastically. For large values of dephasing, the transport gets suppressed again, due to the quantum Zeno effect \cite{rmk09}.

One of the more popular approaches to introducing dissipative effects into PBC models is the addition of anti-Hermitian terms to an otherwise Hermitian Hamiltonian \cite{le96,pp01,mrl08,rmk09,nbb12}.
Such terms can appear, \textit{e.g.}, in a way similar to the Feshbach projection mechanism 
in nuclear physics \cite{fe58}, applications to open quantum systems can be found in Refs. \cite{fbook,ro09,nbb12}.
Alternatively, they can be introduced on phenomenological grounds \cite{le96,pp01,mrl08,rmk09}.
We begin with the exciton model where various protein
pigments are represented by sites labeled by indexes $n,\, m,$ \textit{etc}.
These pigments are coupled to one other by interactions $V_{n m}$, so
the total Hamiltonian can be formally assumed to be of a tight-binding type,
$
\hat{\cal H}
=
\sum\limits_{n}
E_n \ket n \bra n
+
\frac{1}{2}
\sum\limits_{n \not= m}
V_{n m} (\ket n \bra m + \ket m \bra n)
,
$
where the index $1$ labels the donor state,
index $2$ does the acceptor state;
$E_n$ is the energy of a $n$th state.
The total Hilbert space of this system can be divided into two orthogonal subspaces generated by two projection operators, where one subspace is associated with donor-acceptor levels 
and the other is related to the environment.
The basic phenomenological model of EET processes includes two protein co-factors, donor and acceptor, with discrete energy levels, and a third protein pigment which functions as a reservoir.
Using the Feshbach projection method,
one obtains an effective non-Hermitian Hamiltonian that describes the donor-acceptor subsystem
(we use the units $\hbar =1$):
\be\lb{e:ham}
\hat H 
=
\frac{V}{2} \hat\sigma_1
+
\frac{1}{2} (\varepsilon + i \Gamma) \hat\sigma_3
-
\frac{i}{2}  \Gamma \hat I
=
\frac{1}{2}
\twomat{\varepsilon}{V}{V}{-\varepsilon - 2 i \Gamma}
,
\ee
where $\hat\sigma$'s are Pauli matrices, $\hat I$ is the  $2\times 2$ unit matrix,
and
$\varepsilon$ is the difference between renormalized energy levels \cite{nbb12}.
Here
$\Gamma$ is a constant parameter, which effectively describes the cumulative averaged
effect 
of the environment's degrees of freedom \cite{sz13,zs14,sz14cor,sz15,zlo15}. 
In general, $\Gamma$ is a phenomenological parameter, 
and in some special cases, when it is positive,
it can play the role of the decay rate constant.
Under conditions of a weakly-coupled environment one can assume that
\be\lb{e:paramapx}
|\Gamma | \ll |V| < \varepsilon  
.
\ee
For instance, for the quinon-type photosystems 
$\varepsilon \sim 60 \ips$,
$|V| \sim 20 \ips$
and 
$|\Gamma | \sim 1 \ips $ \cite{vm99,la02}.

The time evolution of such a (sub)system is described in general by
a reduced density operator that allows us to consider not only pure states but also mixed ones,
which is important when dealing with open quantum systems \cite{zlo15}.
However, in a theory with a non-Hermitian Hamiltonian, the definition of the statistical 
density operator 
depends on additional physical considerations.

If one assumes that the Hamiltonian (\ref{e:ham})
describes an excitonic (sub)system which is not protected against decay, 
then
the reduced density matrix is defined as a solution
of the following evolution equation:
\be\label{e:eomom}
\frac{d}{dt}
\densnnorm
=-\frac{i}{\hbar}\left[\hat{H}_+,\densnnorm\right]
-\frac{1}{\hbar}\left\{\hat{\Gamma},\densnnorm\right\}
,
\ee
where the square and curly brackets 
denote, respectively, commutator and anti-commutator \cite{fbook}.
Here we have introduced the following Hermitian operators:
$
\hamto_+
=
\frac{1}{2}
(
\hamto + \hamto^\dagger
)
=
\frac{1}{2}
\twomatsm{\varepsilon}{V}{V}{-\varepsilon}
$
and the decay operator
$
\hat\Gamma 
=
\frac{i}{2}
(
\hamto - \hamto^\dagger
)
= 
\twomatsm{0}{0}{0}{\Gamma}
,
$
which correspond to the Hermitian and anti-Her-mitian parts of the Hamiltonian (\ref{e:ham}),
which thus can be written as
$\hamto = \hamto_+ - i \hat\Gamma$.
The eigenvalues of the operator $\hamto_+$ yield energy levels of the excitonic subsystem
in absence of an environment: $(\hamto_+)_j = (-1)^{j+1} \om_0/2 $, where 
$\om_0 = \sqrt{\varepsilon^2 + V^2}$ is the generalized Rabi frequency
and indices $j=1$ and $2$ label the donor and acceptor levels, respectively.

One can easily verify that the trace of the density operator $\densnnorm$ is not conserved
during evolution \cite{sz13}.
In our case it indicates that both donor and acceptor levels will eventually be 
completely depleted, usually at an exponential rate, so that the subsystem 
disappears very fast, within a few picoseconds.
However, this is not what usually happens in reality: 
donor-acceptor systems can hardly become completely ``drained'',
since a level's vacancy would be promptly
occupied by an external particle or quasi-particle.
Moreover, quantum photobiological complexes are able to sustain the transfer of
very large amount of excitons, therefore, their exponentially-fast
disappearance does not seem to be a feature which is pertinent to all possible cases and configurations. 

It turns out that one can account for this phenomenon of sustainability without any modifications of a Hamiltonian:
one merely has to regard
the operator,
\be\lb{e:norm}
\hat\rho = \densnnorm / {\rm tr}\, \densnnorm
,
\ee
as the statistical density operator, instead of $\densnnorm$ 
\cite{zs14,zlo15}.
If, for a given Hamiltonian and initial conditions, 
such an operator does exist and does not exhibit any unphysical properties, then we call
the evolution of such a (sub)system quantum sustainable, 
borrowing the terminology of ref. \cite{z16prb}.
Otherwise, the role of the statistical density operator is played by $\densnnorm$,
and the corresponding evolution would be a non-sustainable process.
Here, sustainability is understood as an ability of a system 
to conserve its
entire probability sample space measure, which is directly related to maintaining the system's 
integrity and separability from its environment
(the term system can refer not only 
to a physical object but also to a process, such as the excitonic energy transfer).
It should be noted that the existence of two types of evolution for the same Hamiltonian, depending on different initial
and boundary conditions defined by physical configuration, is a distinctive feature of the non-Hermitian Hamiltonian approach.
Besides, unlike the also popular (Gorini-Kossakowski-Sudarshan-)Lindblad approach, one does not have to assume
that
environment-induced corrections must obey 
quantum dynamical semigroup symmetry (the latter enforces
the conservation of a density operator's trace and thus cannot account for the above-mentioned effects). 
However, both approaches can be used together, which results 
in 
an ability to describe a wider range of
environmental effects \cite{zs14}.

In principle, in the equation (\ref{e:eomom})
one can change from $\densnnorm$ to $\hat\rho$
and obtain the equation for
the normalized density operator itself
$
\frac{d}{dt}
\hat\rho
=
-\frac{i}{\hbar}
\left[\hat{H}_+,\hat\rho
\right]
-
\frac{1}{\hbar}
\left\{\hat{\Gamma},\hat\rho 
\right\}
+\frac{2}{\hbar} 
\langle \Gamma \rangle
\hat\rho
,
$
where we use the standard notation for quantum-statistical averages $\langle A \rangle = 
{\rm tr} (\hat\rho \, \hat A) = {\rm tr} (\densnnorm \, \hat A)/ {\rm tr} \densnnorm
=
\langle A \rangle'/ {\rm tr}\,\densnnorm$.
Although this equation is difficult to use for the practical purpose of finding solutions, it indicates that the dynamics of the normalized density operator itself is both nonlinear and nonlocal. 
Therefore, the procedure (\ref{e:norm})
despite looking simple by itself, introduces new non-trivial physics. 
As a result, the above-mentioned type of sustainability could lead to quantum-statistical effects which,
while not built-in to the Hamiltonian itself, emerge during the Liouvillian-type evolution of the density operator, as it will be shown in what follows.


\begin{figure}
\centering
\subfloat
{
  \includegraphics[width=\sc\columnwidth]{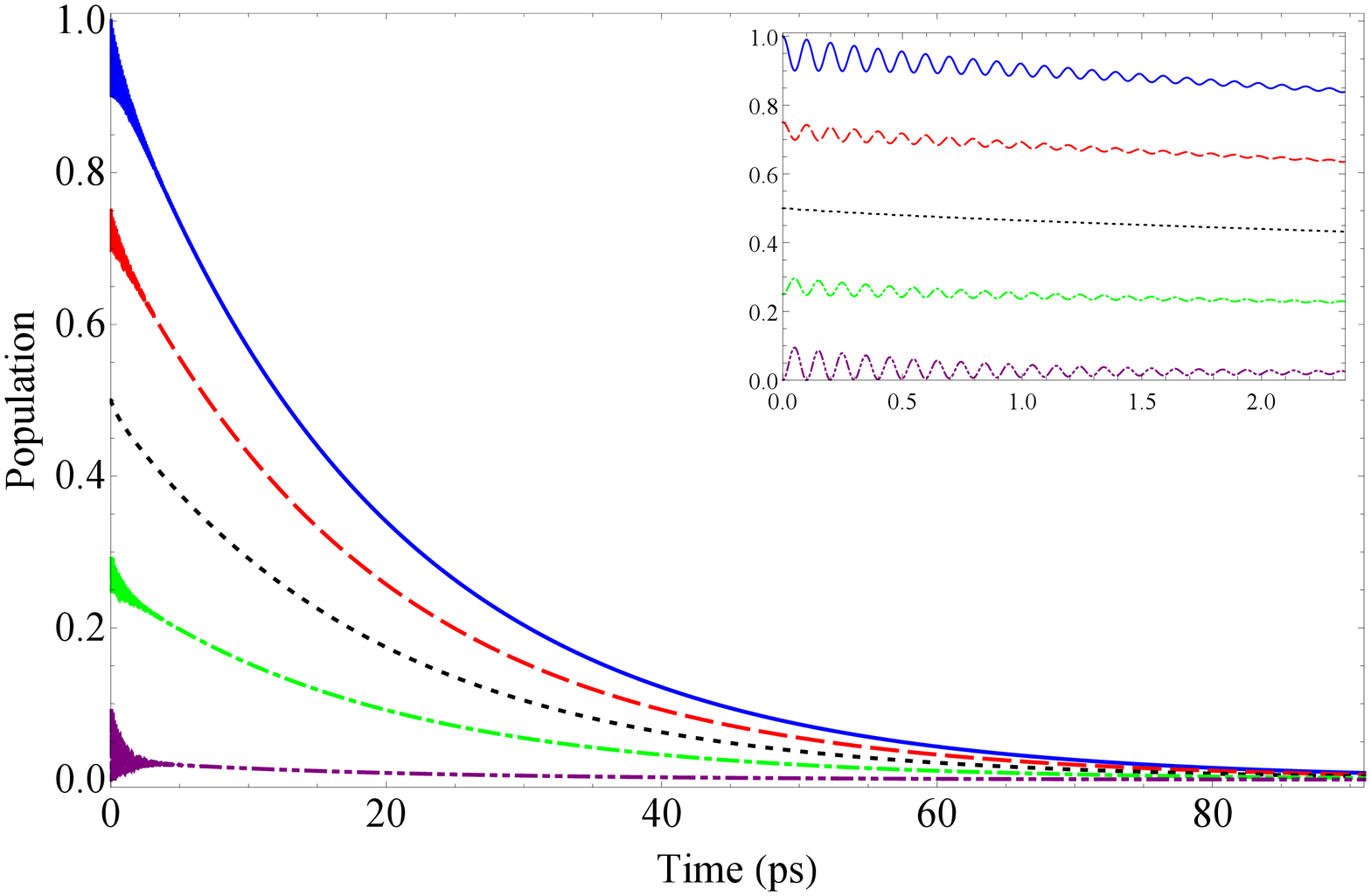}
}
\hspace{0mm}
\subfloat
{
  \includegraphics[width=\sc\columnwidth]{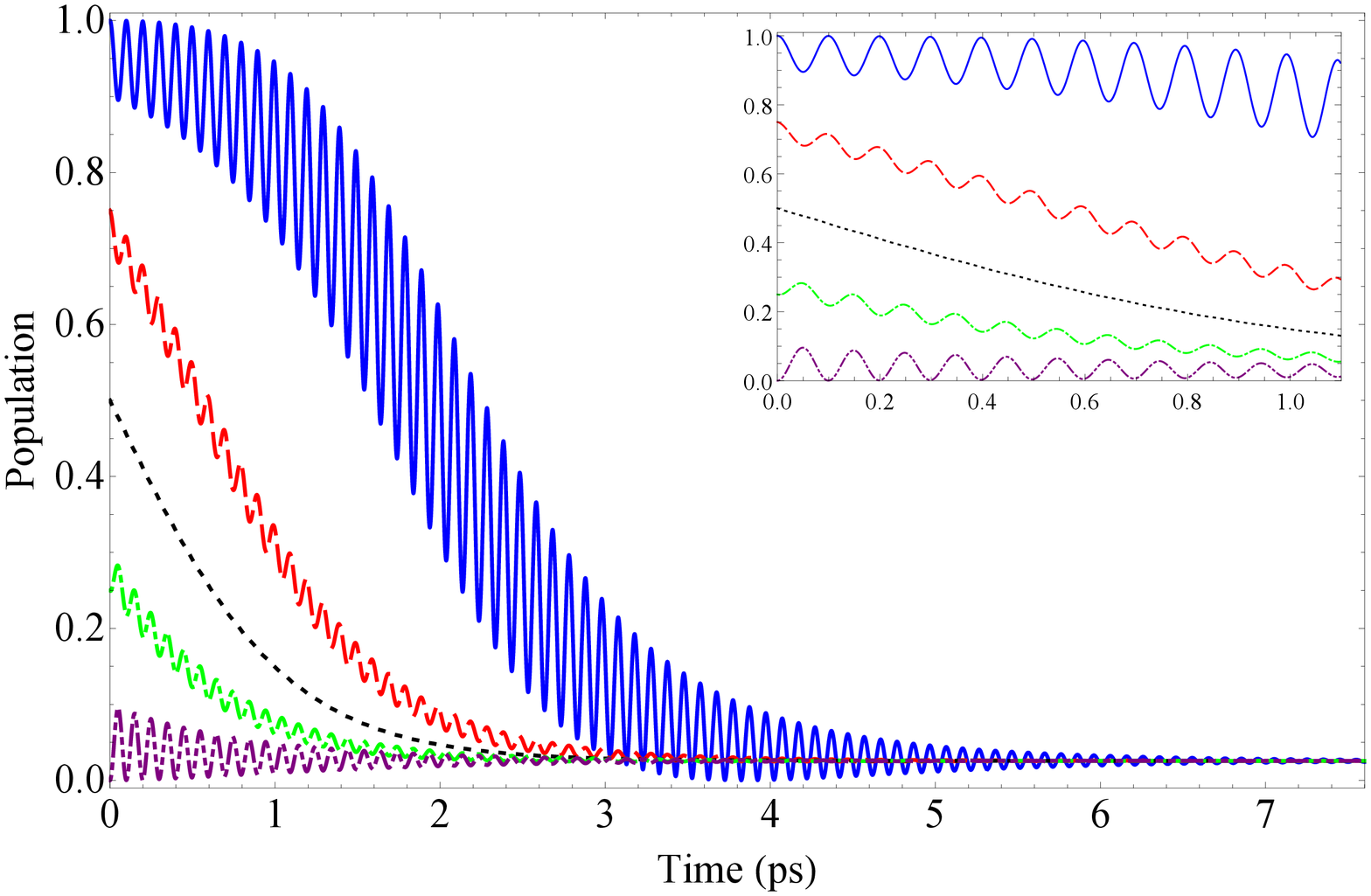}
}
\caption{Comparison of donor level populations $\popdo$
for non-sustainable (upper panel) and sustainable (lower panel) types of evolution,
at various initial conditions.
The excitonic parameters are $\varepsilon = 60 \ips$ and $V = 20 \ips$,
the value of NH parameter $\Gamma$ is $1$ for upper panel and $-1$ for lower panel, 
the values of $p$ are: 
$1$ (solid curves),
$3/4$ (dashed curves), 
$1/2$ (dotted curves),
$1/4$ (dash-dotted curves), 
and 
$0$ (dash-double-dotted curves).
}
\label{f:popdo}
\end{figure}


\begin{figure}
\centering
\subfloat
{
  \includegraphics[width=\sc\columnwidth]{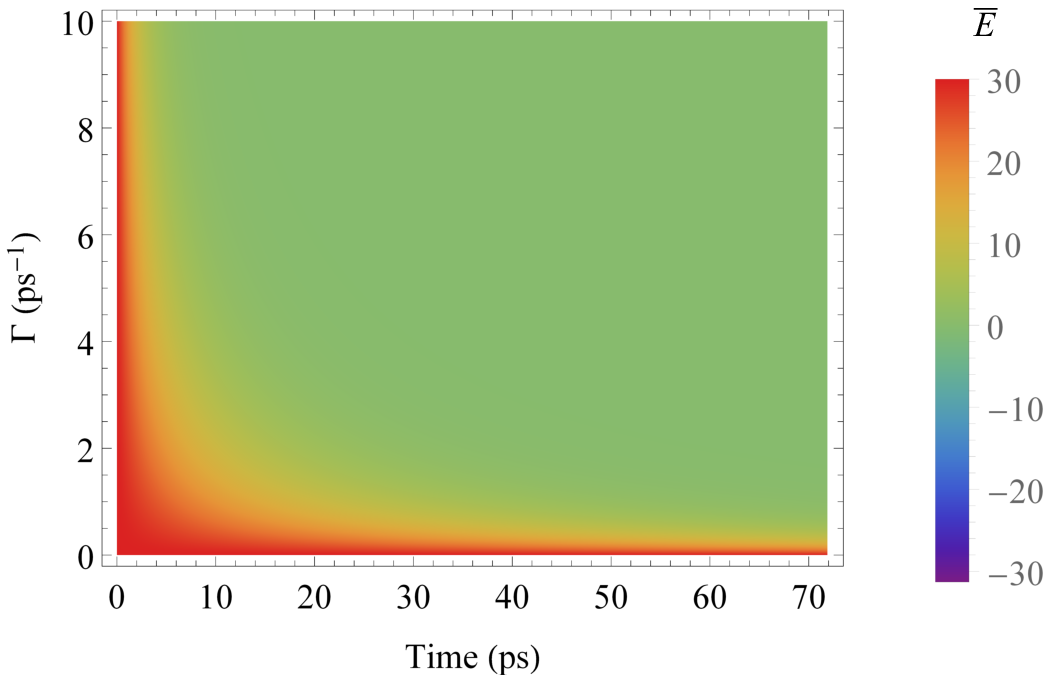}
}
\hspace{0mm}
\subfloat
{
  \includegraphics[width=\sc\columnwidth]{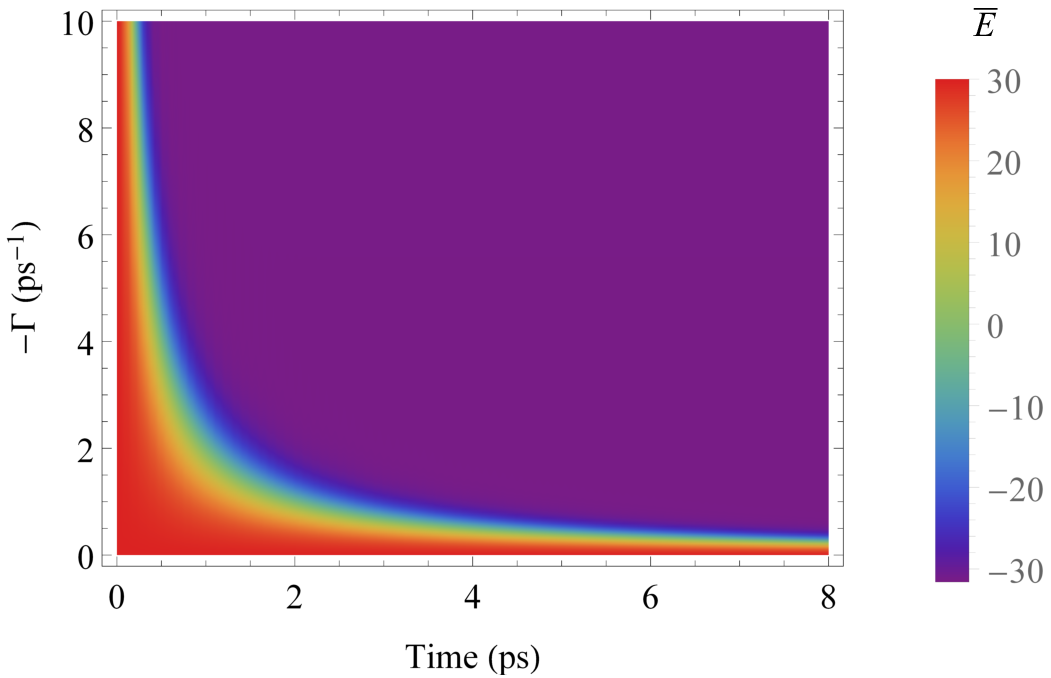}
}
\caption{
Comparison of average energy $\ave$ (in $\ips$)
for non-sustainable (upper panel) and sustainable (lower panel) types of evolution,
at different values of the NH parameter.
The excitonic parameters are $\varepsilon = 60 \ips$ and $V = 20 \ips$,
the initial condition's parameter is $p = 1$.
The deepest red and violet colors correspond to, respectively, donor and acceptor energy
levels in absence of environment.
}
\label{f:encol}
\end{figure}

\begin{figure}
\centering
\subfloat
{
  \includegraphics[width=\sc\columnwidth]{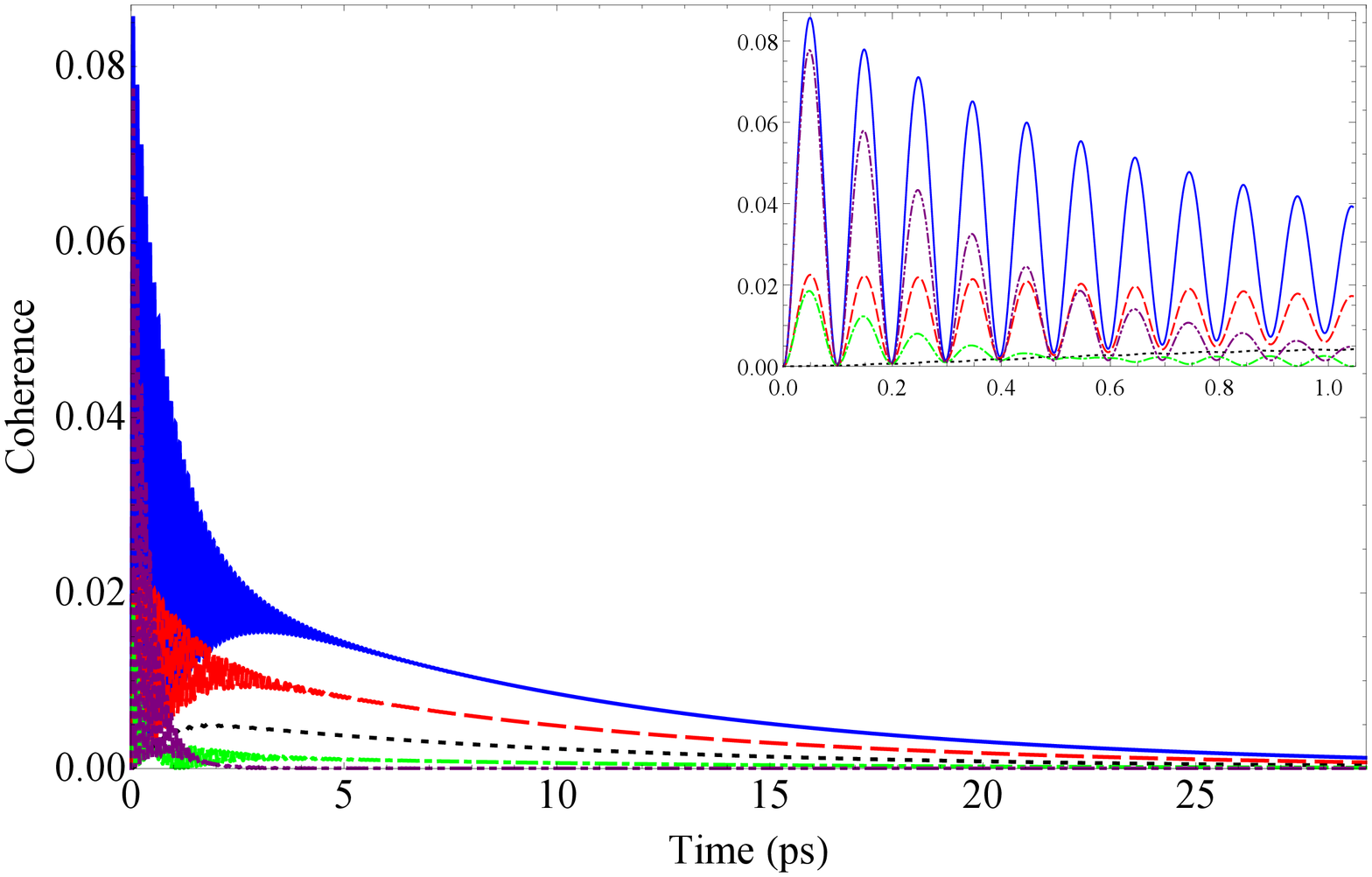}
}
\hspace{0mm}
\subfloat
{
  \includegraphics[width=\sc\columnwidth]{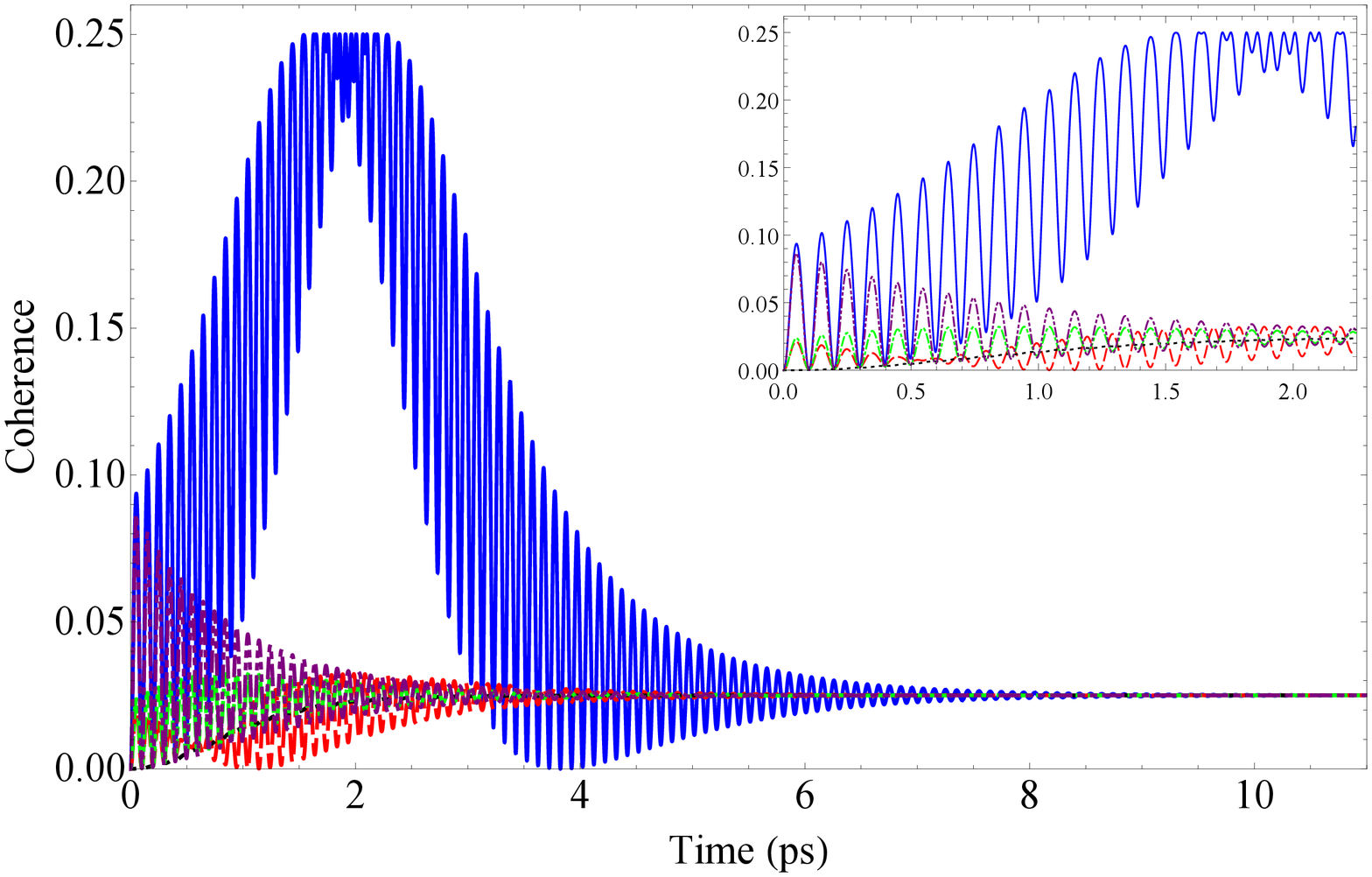}
}
\caption{Comparison of quantum coherence $C$
for non-sustainable (upper panel) and sustainable (lower panel) types of evolution,
at various initial conditions.
The excitonic parameters are $\varepsilon = 60 \ips$ and $V = 20 \ips$,
the value of NH parameter is $1$ for upper panel and $-1$ for lower panel, 
the values of $p$ are: 
$1$ (solid curves),
$3/4$ (dashed curves), 
$1/2$ (dotted curves),
$1/4$ (dash-dotted curves), 
and 
$0$ (dash-double-dotted curves).
}
\label{f:qcoh}
\end{figure}

\begin{figure}
\centering
\subfloat
{
  \includegraphics[width=\sc\columnwidth]{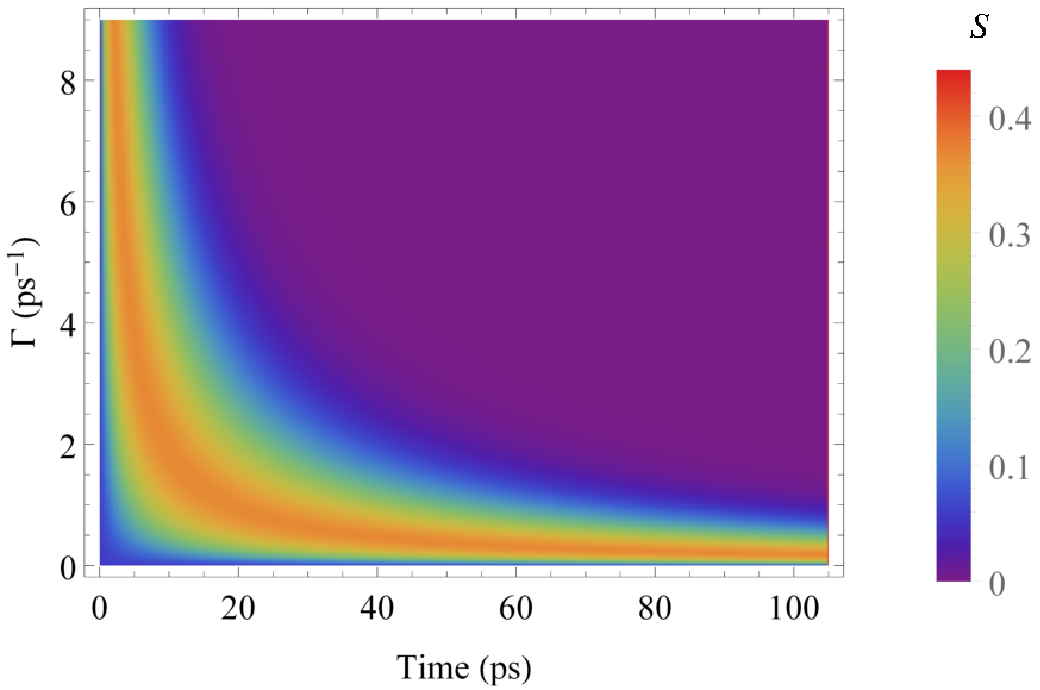}
}
\hspace{0mm}
\subfloat
{
  \includegraphics[width=\sc\columnwidth]{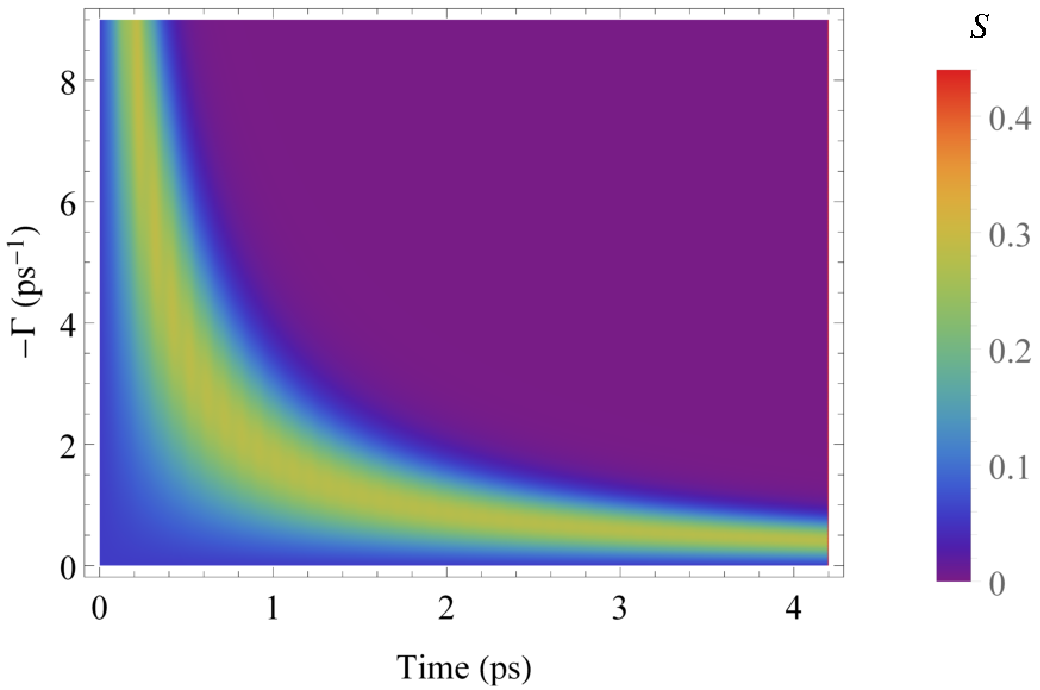}
}
\caption{
Comparison of quantum entropy $S$ (in units of $k_B$) for non-sustainable (upper panel) and sustainable (lower panel) types of evolution,
at different values of the NH parameter.
The excitonic parameters $\varepsilon = 60 \ips$ and $V = 20 \ips$,
the initial condition parameter $p=0.99 \lessapprox 1$ is chosen for plotting here because
at $p=1$ (when an initial state is a donor state exactly) the entropy of the
sustainable case would be identically zero, at any values of other parameters.
}
\label{f:entrcol}
\end{figure}

Assuming initial condition $\densnnorm (0) = \hat\rho (0) =\twomatsm{p}{0}{0}{1-p}$,
$0 \leqslant p \leqslant 1$,
the exact solution of Eq. (\ref{e:eomom}) can be found in the form
$
\densnnorm 
=
\frac{1}{2}
\hat I
{\rm tr} \densnnorm  
+ 
\frac{1}{2}
\sum\limits_{a = 1}^3 \avo{\sigma_a} \hat\sigma_a
$,
where the averages are given by a set
of differential
equations
\be\lb{e:avoeqs}
\frac{d}{d \t}
\fourcol{
\avo{\sigma_1}}{
\avo{\sigma_2}}{
\avo{\sigma_3}}{
\text{tr} \densnnorm
}
=
\begin{pmatrix}
- \Gamma & -\varepsilon & 0 & 0 \\[-0.1ex]
\varepsilon & - \Gamma & -V & 0 \\[-0.1ex]
0 &  V & - \Gamma & \Gamma \\[-0.1ex] 
0 &  0 & \Gamma & - \Gamma 
\end{pmatrix}
\!
\fourcol{
\avo{\sigma_1}}{
\avo{\sigma_2}}{
\avo{\sigma_3}}{
\text{tr}\densnnorm
}
.
\ee
Its solution can be written in a matrix form
as
\be\lb{e:avomsol}
\fourcol{
\avo{\sigma_1}}{
\avo{\sigma_2}}{
\avo{\sigma_3}}{
\text{tr}\densnnorm
}
=
\text{e}^{-\Gamma t}
\textbf{K}
\,
\fourcol{
\cosh{(\om_+ \t)}}{
\sinh{(\om_+ \t)}}{
\cos{(\om_- \t)}}{
\sin{(\om_- \t)}
}
,
\ee
where 
$\kappa = \sqrt{V^4 - 2 V^2 (\Gamma^2 - \varepsilon^2) + (\Gamma^2 +\varepsilon^2)^2}$,
$\om_\pm = $\\ $\sqrt{\kappa \pm (\Gamma^2 - \varepsilon^2 - V^2)}/\sqrt{2}$,
and the coefficient matrix is
\[
\textbf{K}
=
\begin{pmatrix}
\frac{\tilde p  \varepsilon V}{\kappa} & \frac{\varepsilon V \Gamma}{\kappa \om_+}& - \frac{\tilde p  \varepsilon V }{\kappa}& -\frac{\varepsilon V \Gamma}{\kappa \om_-} \\
- \frac{V \Gamma}{\kappa} & - \frac{\tilde p \om_+ V}{\kappa} & \frac{V \Gamma}{\kappa}  & \frac{\tilde p  V (\om_+^2 - \kappa)}{\kappa \om_-} \\
\frac{\tilde p (\om_+^2 + \varepsilon^2) }{\kappa}&  \frac{\Gamma (\om_+^2 + \varepsilon^2)}{\kappa \om_+} & 
\frac{\tilde p  (\kappa - \om_+^2 - \varepsilon^2)}{\kappa} & \frac{\Gamma  (\kappa - \om_+^2 - \varepsilon^2)}{\kappa \om_-}  \\
\frac{\kappa - \om_+^2 + \Gamma^2 }{\kappa} &  \frac{\tilde p \Gamma(\om_+^2 + \varepsilon^2)}{\kappa \om_+} & 
\frac{\om_+^2 - \Gamma^2}{\kappa} & \frac{\tilde p \Gamma  (\kappa - \om_+^2 - \varepsilon^2)}{\kappa \om_-} 
\end{pmatrix}
,
\]
where $\tilde p = 2 p -1$ (if an initial density matrix corresponds to the donor state then $p=\tilde p = 1$).
In the approximation of a weakly-coupled environment (\ref{e:paramapx})
the main parameters simplify to
$
\om_+
\approx
\frac{\varepsilon |\Gamma |}{\om_0}$,
$
\om_-
\approx
\om_0 
-\frac{V^2 \Gamma^2}{2 \om_0^3}
$
,
$
\kappa
\approx
\om_0^2 
+
\frac{\Gamma^2 (\varepsilon^2-V^2)}{\om_0^2}
.$
As for the normalized density operator, it is algebraically given by Eq. (\ref{e:norm}),
which yields:
$
\hat\rho 
=
\frac{1}{2}
\hat I
+ 
\frac{1}{2}
\sum\limits_{a = 1}^3 \av{\sigma_a} \hat\sigma_a
,
$
where $\av{\sigma_a} = \avo{\sigma_a} / {\rm tr} \densnnorm $.

With solution (\ref{e:avomsol}) in hands, one can easily compute the following observables:
population of the donor level
$$
\popdo
=
\left\{
\baa{l}
(\densnnorm)_{11} = \frac{1}{2} ({\rm tr} \densnnorm + \avo{\sigma_3})
\\[0.5ex]
(\hat\rho)_{11} = \frac{1}{2} (1+\av{\sigma_3})
\eaa
\right.
,
$$
average energy
$$
\ave
=
\left\{
\baa{l}
\text{tr}(\densnnorm \hamto_+)
\\[0.5ex]
\text{tr}(\hat\rho \hamto_+)
\eaa
\right.
,$$
measure of coherence
$$
C
=
\left\{
\baa{l}
|(\densnnorm)_{12}|^2 
= 
\frac{1}{4} (\avo{\sigma_1}^2 + \avo{\sigma_2}^2)
\\[0.5ex]
|(\hat\rho)_{12}|^2 
= 
\frac{1}{4} (\av{\sigma_1}^2 + \av{\sigma_2}^2)
\eaa
\right.
,
$$
and
Gibbs-von-Neumann
entropy 
$$
S
=
\left\{
\baa{l}
-k_B \text{tr}(\densnnorm \ln{\densnnorm})
\\[0.5ex]
-k_B \text{tr}(\hat\rho \ln{\hat\rho})
\eaa
\right.
,
$$
where
upper and lower cases refer, respectively, to non-sustainable and sustainable types of evolution, \textit{i.e.}, to those described by non-normalized and normalized density operators.

In the approximation of weakly-coupled environment (\ref{e:paramapx}),
the observables have 
the following large-time asymptotic behavior:
\ba&&
\popdo \to
\left\{
\baa{l}
(2 p k_+ \om_0^2 - \tilde p V^2)
(2\om_0)^{-2}
\text{exp}(k_- \Gamma t)
, 
\\[0.5ex]
k_+ /2
, 
\eaa
\right.
\nn\\
&&
\ave 
\to
\left\{
\baa{l}
\left[
(2 p + k_-) \om_0^2
- \tilde p V^2
\right]
(4 \varepsilon)^{-1}
\,
\text{exp}(k_- \Gamma t)
, 
\\[0.5ex]
(k_- + 1) \om_0^2 / (2 \varepsilon )
, 
\eaa
\right.
\lb{e:asym}
\\&&
C \to
\left\{
\baa{l}
(V / 4 \om_0)^2 
(2 p + \tilde p k_-)^2
\,
\text{exp}(2 k_- \Gamma t)
, 
\\[0.5ex]
(V / 2 \om_0)^2
, 
\eaa
\right.
\nn
\ea
where 
$k_\pm = \varepsilon\, \Sign{\Gamma} \om_0^{-1} \pm 1$.

From the evolution of the observables, provided in Figs. \ref{f:popdo}-\ref{f:entrcol},
and their large-time asymptotics (\ref{e:asym}), 
a few important conclusions can be drawn.
First of all, Figs. \ref{f:popdo}-\ref{f:entrcol} reveal that the
environment represented by the NH parameter $\Gamma$ plays a crucial role in the EET process, even if it is weakly coupled to the excitonic system, as defined in Eq. (\ref{e:paramapx}).
In fact, when $\Gamma$ is exactly zero the excitonic system undergoes plain Rabi-type oscillations, and therefore energy transfer is not facilitated. 
Once $\Gamma$ acquires a value, however small, the oscillations become damped, and energy passes through the system more easily. 
Additionally, the possibility of sustainable evolution expands the admissible parameter space of the NH models containing the parameter $\Gamma$ --
the latter
now can be both positive and negative.
Secondly, Figs. \ref{f:popdo}-\ref{f:encol} demonstrate that, at the same absolute value of the NH parameter, the discharge of the donor level (and therefore the transfer of energy through the system) occurs much faster for sustainable evolution than for non-sustainable, at least by an order of magnitude. 
Despite sustainable evolution preserving a small residual population of donor level at large times (to prevent the excitonic subsystem from complete disappearance) the majority of energy is transferred through the system. The physical interpretation of this residual population will be further discussed below. 
Thirdly, Fig. 
\ref{f:encol} shows that the average energy for sustainable evolution tends to the acceptor level $(\hamto_+)_2$ at large times,
whereas for non-sustainable evolution energy discharge stops halfway to the acceptor level. Therefore, regardless of the NH parameter's value (as long as it is not zero), the discharge of the donor level is more complete for sustainable evolution than for non-sustainable. As a result, sustainable open excitonic systems are capable of transferring larger portions of energy per system than non-sustainable ones.
Furthermore, 
from Fig. \ref{f:qcoh} and analytical formulae for large-time behavior of $C$, one can deduce that for non-sustainable evolution, quantum coherence vanishes exponentially at large times, whereas for sustainable evolution it tends to a small constant, approximately $(V / 2 \om_0)^2$.
This indicates that quantum coherence, after having an initial beat followed by a rapid decrease to a small residual value, would remain non-zero in sustainable photobiological systems for considerably longer. This residual coherence, together with the above-mentioned residual donor level population, points to the appearance of a metastable state which significantly slows down the dephasing processes in PBCs. 
Besides, in an ensemble consisting of many excitonic systems, such behavior leads to quantum beating between different exciton level groups.
Finally,
Fig. \ref{f:entrcol} shows that for sustainable systems, at the same absolute value of the NH parameter, quantum entropy has smaller peak values and vanishes considerably faster than for non-sustainable ones. Therefore, the emergence of sustainable evolution in the models described by NH Hamiltonians explains why photobiological systems are durable and resistant to external dissipative effects.







\noindent
\textbf{Acknowledgments}. 
This work is based on the research supported by the National Research Foundation of South Africa.
Proofreading of the manuscript by P. Stannard is greatly appreciated.


\begin{thebibliography}{0}



\bibitem{bm75}
R. E. Fenna \andd B. W. Matthews, 
\jrn{R. E. Fenna \andd B. W. Matthews}{Chlorophyll arrangement in a bacteriochlorophyll protein from Chlorobium limicola}{Nature}{258}{573-577}{1975}

\bibitem{lzb97}
Y.-F. Li, 
\textit{et al.},
\jrn{Y.-F. Li, W. Zhou, R. E. Blankenship, \andd J. P. Allen}{Crystal Structure of the Bacteriochlorophyll a Protein
from Chlorobium tepidum}{J. Mol. Biol.}{271}{456-471}{1997}


\bibitem{wgr02}
J. Wang, 
\textit{et al.},
\jrn{J. Wang, D. Gosztola, S. V. Ruffle, C. Hemann, M. Seibert, M. R. Wasielewski, R. Hille, T. L. Gustafson, \andd R. T. Sayre}{Functional asymmetry of photosystem II D1 and D2 peripheral chlorophyll mutants of Chlamydomonas reinhardtii}{Proc. Natl. Acad. Sci. USA}{99}{4091}{2002}

\bibitem{xsg04}
L. Xiong, 
\textit{et al.},
\jrn{L. Xiong, M. Seibert, A. Gusev, M. Wasielewski, C. Hemann, C. R. Hille, \andd R. T. Sayre}{Substitution of a Chlorophyll into the Inactive Branch Pheophytin-Binding Site Impairs Charge Separation in Photosystem II}{J. Phys. Chem. B}{108}{16904}{2004}

\bibitem{bsv05}
T. Brixner, 
\textit{et al.},
\jrn{T. Brixner, J. Stenger, H. M. Vaswani, M. Cho, R. E. Blankenship \andd G. R. Fleming}{Two-dimensional spectroscopy of electronic couplings in photosynthesis}{Nature}{434}{625-628}{2005}

\bibitem{blbook}
R. E. Blankenship, 
Molecular Mechanisms of Photosynthesis
(Wiley-Blackwell, New Jersey, 2002).



\bibitem{fhg89}
S. E. Freeman, 
\textit{et al.},
\jrn{S. E. Freeman, H. Hacham, R. W. Gange, D. J. Maytum, J. C. Sutherland, \andd B. M. Sutherland}{Wavelength dependence of pyrimidine dimer formation in DNA of human skin irradiated in situ with ultraviolet light}{Proc. Natl. Acad. Sci. USA}{86}{5605-5609}{1989}

\bibitem{wpc01}
S. E. Whitmore, 
\textit{et al.},
\jrn{S. E. Whitmore, C. S. Potten, C. A. Chadwick, P. T. Strickland, \andd W. L. Morison}{Effect of photoreactivating light on UV radiation-induced alterations in human skin}{Photodermatol. Photoimmunol. Photomed.}{17}{213-217}{2001}


\bibitem{ay05}
N. Agar \andd A. R. Young, 
\jrn{N. Agar \andd A. R. Young}{Melanogenesis: a photoprotective response to DNA damage?}{Mutation Research: Fundam. Mol. Mech. Mutagen.}{571}{121-132}{2005}

\bibitem{pwm15}
S. Premi, 
\textit{et al.},
\jrn{S. Premi, S. Wallisch, C. M. Mano, A. B. Weiner, A. Bacchiocchi, K. Wakamatsu, 
E. J. H. Bechara, R. Halaban, T. Douki, \andd D. E. Brash}{Chemiexcitation of melanin derivatives induces DNA photoproducts long after UV exposure}{Science}{347}{842-847}{2015}





\bibitem{sbs97}
S. Savikhin, D. R. Buck, \andd W. S. Struve, 
\jrn{S. Savikhin, D. R. Buck, \andd W. S. Struve}{Oscillating anisotropies in a bacteriochlorophyll
protein: Evidence for quantum beating between exciton levels}{Chem. Phys.}{223}{303-312}{1997}

\bibitem{ecr07}
G. S. Engel, 
\textit{et al.},
\jrn{G. S. Engel, T. R. Calhoun, E. L. Read, T.-K. Ahn, T. Man\v cal, Y.-C. Cheng, R. E. Blankenship \andd G. R. Fleming}{Evidence for wavelike energy transfer through quantum coherence in photosynthetic systems}{Nature}{446}{782-786}{2007} 

\bibitem{lcf07}
H. Lee, Y.-C. Cheng, \andd G. R. Fleming, 
\jrn{H. Lee, Y.-C. Cheng, \andd G. R. Fleming}{Coherence Dynamics in Photosynthesis: Protein Protection of Excitonic Coherence}{Science}{316}{1462-1465}{2007} 

\bibitem{ran14}
E. Romero, 
\textit{et al.},
\jrn{E. Romero, R. Augulis, V. I. Novoderezhkin, M. Ferretti, J. Thieme, D. Zigmantas \andd R. van Grondelle}{Quantum coherence in photosynthesis for efficient solar-energy conversion}{Nature Phys.}{10}{676-682}{2014}



\bibitem{cs84}
V. \v C\'apek \andd V. Sz\"ocs, 
\jrn{V. \v C\'apek \andd V. Sz\"ocs}{Is the Sink Model of Exciton Trapping in Molecular Condensates Satisfactory?}{Phys. Status Solidi B}{125}{K137-K142}{1984}

\bibitem{fbook}
F. H. M. Faisal,  
Theory of Multiphoton Processes (Plenum Press, New York, 1986). 


\bibitem{le96}
J. A. Leegwater, 
\jrn{J. A. Leegwater}{Coherent versus Incoherent Energy Transfer and Trapping in Photosynthetic Antenna
Complexes}{J. Phys. Chem.}{100}{14403-14409}{1996}

\bibitem{pp01}
R. Pin\v c\'ak \andd M. Pudlak, 
\jrn{R. Pin\v c\'ak \andd M. Pudlak}{Noise breaking the twofold symmetry of photosynthetic reaction centers: Electron transfer}{Phys. Rev. E}{64}{031906}{2001}

\bibitem{mrl08}
M. Mohseni,
\textit{et al.},
\jrn{Mohseni M, Rebentrost P, Lloyd S \andd Aspuru-Guzik A}{Environment-assisted quantum walks in photosynthetic energy transfer}{J. Chem. Phys.}{129}{174106}{2008}


\bibitem{oc08}
A. Olaya-Castro,
\textit{et al.},
\jrn{Olaya-Castro A, Chiu Fan Lee, Francesca Fassioli Olsen, \andd Neil F. Johnson}{Efficiency of energy transfer in a light-harvesting system under quantum coherence}{Phys. Rev. B}{78}{085115}{2008}


\bibitem{ph08}
M.B. Plenio \andd S.F. Huelga,
\jrn{M.B. Plenio \andd S.F. Huelga}{Dephasing assisted transport: Quantum networks and biomolecules}{New J. Phys.}{10}{113019}{2008}


\bibitem{if09}
A. Ishizaki \andd G. R. Fleming, 
\jrn{A. Ishizaki \andd G. R. Fleming}{Theoretical examination of quantum coherence in a photosynthetic system at physiological temperature}{Proc. Natl. Acad. Sci. USA}{106}{17255}{2009}



\bibitem{rmk09}
P. Rebentrost, 
\textit{et al.},
\jrn{P. Rebentrost, M. Mohseni, I. Kassal, S. Lloyd, \andd A. Aspuru-Guzik}{Theoretical examination of quantum coherence in a photosynthetic system at physiological temperature}{New J. Phys.}{11}{033003}{2009}


\bibitem{cdc10}
A. W. Chin, 
\textit{et al.},
\jrn{A. W. Chin, A. Datta, F. Caruso, S. F. Huelga, \andd M. B. Plenio}{Noise-assisted energy transfer in quantum networks and light-harvesting complexes}{New J. Phys.}{12}{065002}{2010}


\bibitem{nbb12}
A. I. Nesterov, G. P. Berman, \andd A. R. Bishop, 
\jrn{A. I. Nesterov, G. P. Berman, \andd A. R. Bishop}{Non-Hermitian approach for modeling of noise-assisted
quantum electron transfer in photosynthetic complexes}{Fortschr. Phys.\!}{\!}{1-16}{2012}

\bibitem{nmm15}
P. Nalbach, C. A. Mujica-Martinez, \andd M. Thorwart, 
\jrn{P. Nalbach, C. A. Mujica-Martinez, \andd M. Thorwart}{Vibronically coherent speed-up of the excitation energy transfer in the Fenna-Matthews-Olson complex}{Phys. Rev. E}{91}{022706}{2015}

\bibitem{mmn15}
C. A. Mujica-Martinez \andd P. Nalbach, 
\jrn{C. A. Mujica-Martinez \andd P. Nalbach}{On the influence of underdamped vibrations on coherence
and energy transfer times in light-harvesting complexes}{Ann. Phys. (Berlin)}{527}{592-600}{2015}


\bibitem{an58}
P. W. Anderson, 
\jrn{P. W. Anderson}{Absence of Diffusion in Certain Random Lattices}{Phys. Rev.}{109}{1492}{1958}



\bibitem{fe58}
H. Feshbach, 
\jrn{H. Feshbach}{Unified Theory of Nuclear Reactions}{Ann. Phys.}{5}{357-390}{1958}


\bibitem{ro09}
I. Rotter,
\jrn{I. Rotter}{A non-Hermitian Hamilton operator and the physics
of open quantum systems}{J. Phys. A: Math. Theor.}{42}{153001}{2009}



\bibitem{sz13}
A. Sergi \andd K. G. Zloshchastiev, 
\jrn{Sergi, A. \andd Zloshchastiev, K. G.}{Non-Hermitian quantum dynamics of a two-level system and models of dissipative environments}{Int. J. Mod. Phys. B}{27}{1350163}{2013}

\bibitem{zs14}
K. G. Zloshchastiev \andd A. Sergi,
\jrn{Zloshchastiev, K. G. \andd Sergi, A.}{Comparison and unification of non-Hermitian and Lindblad approaches with applications to open quantum optical systems}{J. Mod. Optics}{61}{1298-1308}{2014}

\bibitem{sz14cor}
A. Sergi  and  K. G. Zloshchastiev,
\jrn{Sergi A. \andd Zloshchastiev K. G.}{Time correlation functions for non-Hermitian quantum systems}{Phys. Rev. A}{91}{062108}{2015}

\bibitem{sz15}
A. Sergi and K. G. Zloshchastiev,
J. Stat. Mech. \textbf{2016}, 033102 (2016);\\
A. Sergi and P. V. Giaquinta, Entropy \textbf{18}, 451 (2016).

\bibitem{zlo15}
K. G. Zloshchastiev,
\jrn{Zloshchastiev K. G.}{Non-Hermitian Hamiltonians and stability of pure states in quantum mechanics}{Eur. Phys. J. D}{69}{253}{2015} 



\bibitem{vm99}
M. H. Vos \andd J.-L. Martin,
\jrn{M. H. Vos \andd J.-L. Martin}{Femtosecond processes in proteins}{Biochim. Biophys. Acta: Bioenergetics}{1411}{1-20}{1999}

\bibitem{la02}
V. D. Lakhno,
\jrn{V. D. Lakhno}{Oscillations in the primary charge separation in bacterial
photosynthesis}{Phys. Chem. Chem. Phys.}{4}{2246-2250}{2002}

\bibitem{z16prb}
K. G. Zloshchastiev,
\jrn{Zloshchastiev K. G.}{Quantum-statistical approach to electromagnetic wave propagation and dissipation inside dielectric media and nanophotonic and plasmonic waveguides}{Phys. Rev. B}{94}{115136}{2016}











\end{thebibliography}
\end{document}